\documentclass{PoS}
\usepackage{graphicx}
\usepackage{amsfonts} % AMS
\usepackage{amssymb} % AMS
\usepackage{amsmath} % AMS
\usepackage{caption}

\usepackage[dvipsnames]{xcolor}
\usepackage{pagecolor}
\DeclareMathOperator{\Tr}{Tr}

\graphicspath{{figs/}}

\title{Neutron Electric Dipole Moments with Clover Fermions}

\ShortTitle{Neutron EDM with Clover Fermions}

\author{\speaker{Boram Yoon}\\
        Los Alamos National Laboratory, Computer, Computational, and Statistical Sciences
CCS-7, Los Alamos, NM 87545\\
        E-mail: \email{boram@lanl.gov}}

\author{Tanmoy Bhattacharya\\
        Los Alamos National Laboratory, Theoretical Division
T-2, Los Alamos, NM 87545\\
        E-mail: \email{tanmoy@lanl.gov}}

\author{Vincenzo Cirigliano\\
        Los Alamos National Laboratory, Theoretical Division
T-2, Los Alamos, NM 87545\\
        E-mail: \email{cirigliano@lanl.gov}}

\author{Rajan Gupta\\
        Los Alamos National Laboratory, Theoretical Division
T-2, Los Alamos, NM 87545\\
        E-mail: \email{rg@lanl.gov}}

\abstract{
We present preliminary results for the contributions to the neutron EDM arising from the QCD $\theta$-term, the Weinberg three-gluon and the quark chromo-EDM operators from our ongoing lattice calculations using clover valence quarks on the MILC HISQ lattices. We use the gradient-flow technique to smooth the lattices and renormalize the gluonic operators, and use the Schwinger source method to incorporate the quark chromo-EDM interactions in the quark propagator. For the QCD $\theta$-term and the Weinberg three-gluon operator, we report results in the gradient-flow scheme from 8 ensembles at four lattice spacings and three pion masses, including 2 physical pion mass ensembles described in~Table~\ref{tab:confs}. For the quark chromo-EDM, unrenormalized results are presented at two lattice spacings, $a=0.12$ and $0.09$~fm, and two pion masses, $M_\pi = 310$~MeV and $220$~MeV.
}

\FullConference{The 37th International Symposium on Lattice Field Theory - LATTICE2019\\
		16-22 June 2019\\
		Wuhan, China.}

\begin{document}

\section{Introduction}
Neutrons can have nonvanishing electric dipole moment (EDM) if the theory has broken P and T symmetries, or CP violation (CPV). Since CPV in the standard model (SM) is small or strongly suppressed at high temperature, new CPV from beyond the SM (BSM) is needed to explain matter-antimatter asymmetry via baryogenesis, and EDMs of elementary particles are good probes of it. The CPV interactions of interest in the low-energy effective Lagrangian are of dimension 4--6:\looseness-1
\begin{align}
\mathcal{L}_\text{CPV}^{d=4,5,6} = & 
-\frac{\mathrm{g}^2}{32\pi^2} \bar{\theta} G\tilde{G} 
-\frac{i}{2}\sum_{q=u,d,s}d_q\bar{q}(\sigma_{\mu\nu} F^{\mu\nu})\gamma_5 q
-\frac{i}{2}\sum_{q=u,d,s}\tilde{d}_q \bar{q}(\sigma_{\mu\nu} G^{\mu\nu})\gamma_5 q
\nonumber \\
&+d_w \frac{\mathrm{g}}{6}f^{abc}G^a_{\mu\nu}\tilde{G}^{\nu\rho, b}G_\rho^{\mu, c} + \sum_i C_i^{(4q)}O_i^{(4q)}\,,
\label{eq:L_eff}
\end{align}
where $\tilde{G}^{\mu\nu, b} = \varepsilon^{\mu\nu\alpha\beta}G^b_{\alpha\beta}/2$. Here, the terms on the r.h.s are the QCD $\theta$-term (d=4); quark EDM (qEDM) and quark chromo-EDM (CEDM) (d=5), and the Weinberg's three-gluon operator ($W_{ggg}$) and various four-quark operators (d=6). In this paper, we will discuss the calculation of the neutron EDM induced by the QCD $\theta$, $W_{ggg}$, and the CEDM terms.

\section{Neutron EDM from QCD $\theta$-term and Weinberg's three-gluon operator}
Expectation values of observables in the presence of the $\theta$-term can be calculated using standard lattices generated without the $\theta$-term in the action by exploiting the small-$\theta$ expansion \cite{Shintani:2005xg}\looseness-1
\begin{align}
\label{eq:theta-exp}
  \langle {O}(x)\rangle_{\bar\theta} 
    = \frac{1}{Z_{\bar\theta}} \int d[U,q,\overline{q}]
      {O}(x) e^{-S_{QCD} - i\theta Q}
    = \langle {O}(x)\rangle_{\bar\theta=0} - i\bar\theta\langle {O}(x) Q\rangle_{\bar\theta=0}
      + \mathcal{O}({\bar\theta}^2)\,,
\end{align}
where $Q$ is the topological charge $Q = \int d^4x \frac{G\tilde{G}}{32\pi^2}$. Since the phenomenological estimate, $\theta \lesssim \mathcal{O}(10^{-10})$ \cite{Pospelov:1999mv}, is tiny, the leading order term in $\theta$ suffices. 

\begin{figure}[tbp]
\begin{center}
\includegraphics[width=0.32\textwidth]{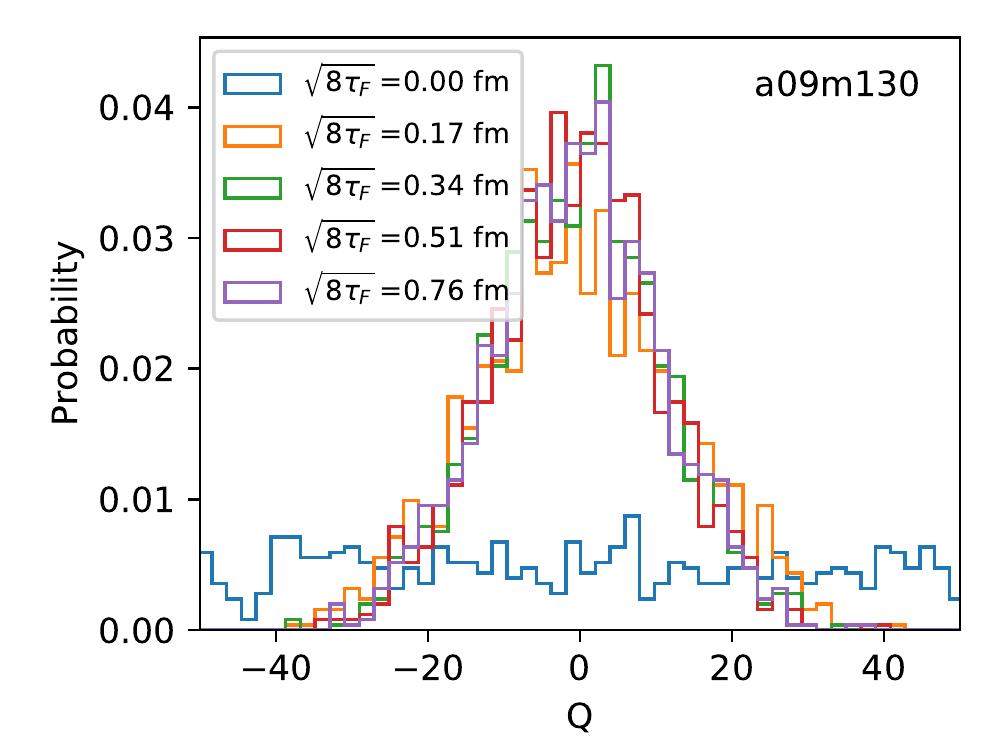}
\includegraphics[width=0.32\textwidth]{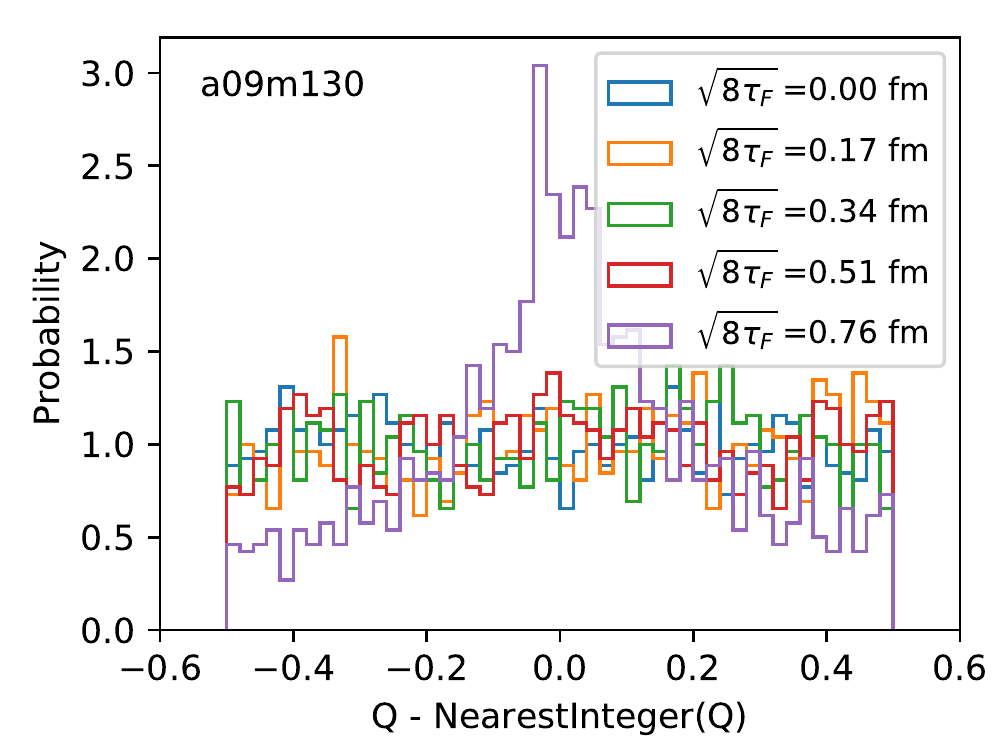}
\includegraphics[width=0.32\textwidth]{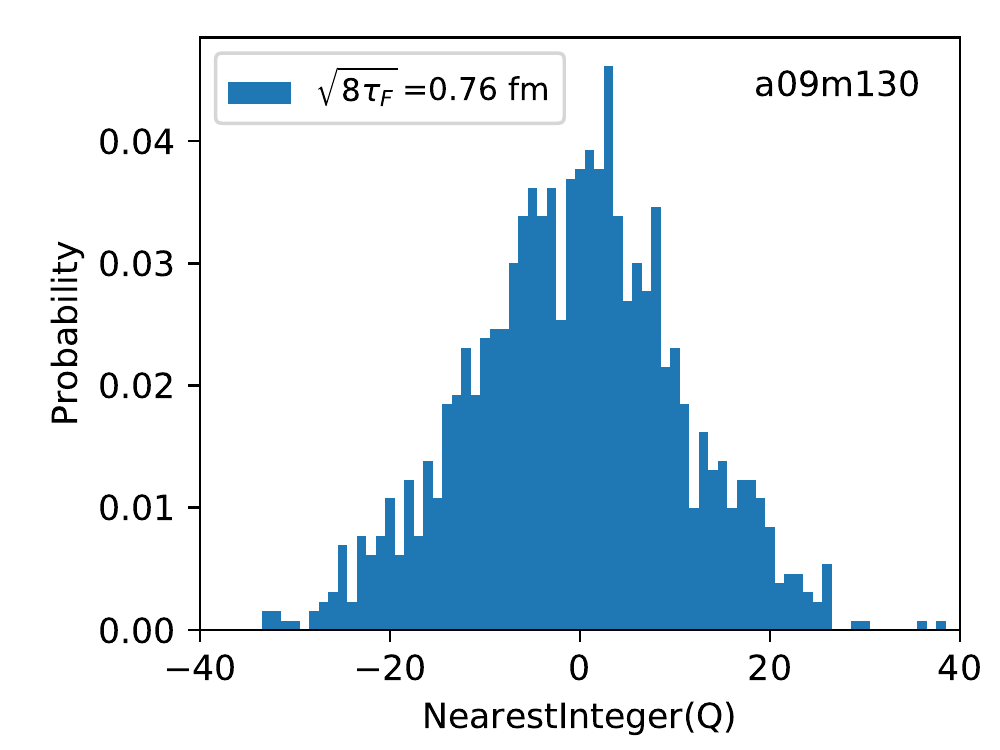}\\
\includegraphics[width=0.32\textwidth]{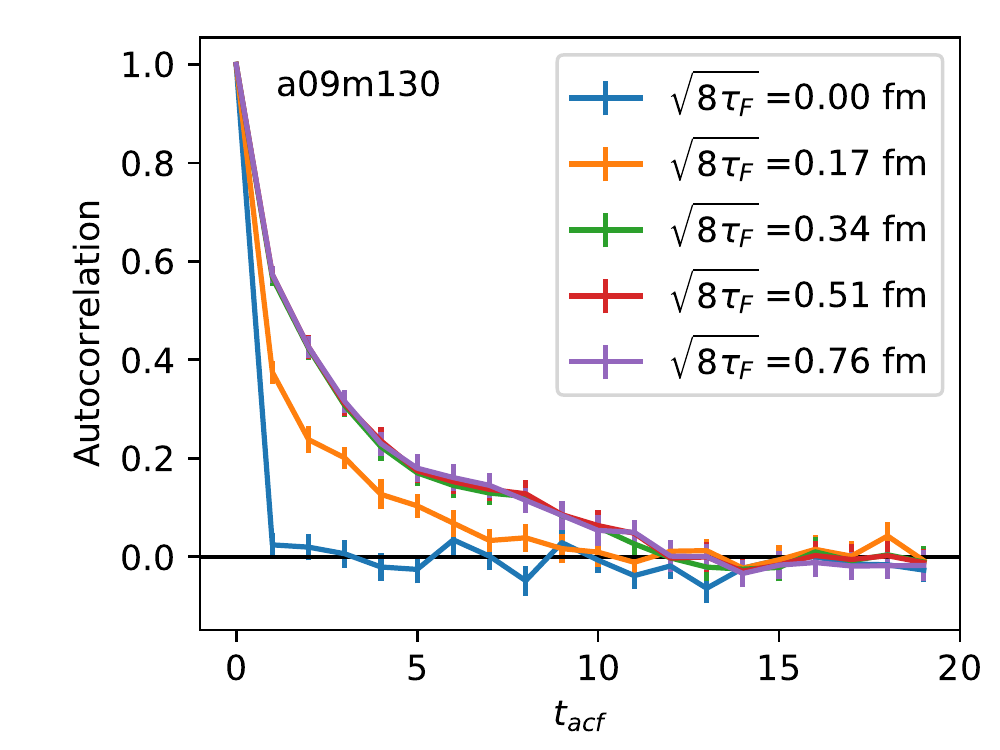}
\includegraphics[width=0.55\textwidth]{{{Qtop_traj_a09m130_0.76}}}
\end{center}
\vspace{-0.5cm}
\caption{Various aspects of the topological charge calculated on the $a09m130$ HISQ ensemble for five different gradient flow times $\tau_F$: (left top) distribution of $Q$, (center top) distribution of the non-integer part of $Q$, (right top) distribution of the integer part of $Q$, (left bottom) autocorrelation function, and (right bottom) Q as a function of the lattice trajectory time.
\label{fig:Q}}
\end{figure}

We calculate the topological charge using the $\mathcal{O}(a^4)$-improved field-strength tensor \cite{BilsonThompson:2002jk} with gradient flow \cite{Luscher:2010iy} on the MILC HISQ lattices \cite{Bazavov:2012xda}.\footnote{Throughout this paper, we will use the notation $aABmXYZ$ to denote an ensemble, where $AB$ represents the approximated lattice spacing in units of $0.01$~fm, and $XYZ$ represents the pion mass in units of MeV.} After analyzing 10 different ensembles with $a=0.15-0.06$~fm and $M_\pi=310-130$~MeV, we find that (i) $Q$ converges to a stable distribution after the gradient flow time $\tau_F \approx 0.34$~fm; however, (ii)  it requires much longer $\tau_F$ for $Q$ to converge to an integer, and this  $\tau_F$ depends on $a$ and $M_\pi$; coarser $a$ or smaller $M_\pi$ lattices need longer $\tau_F$; and (iii) very long autocorrelations length longer than 30 configurations are found in the $a06m310$ and $a06m220$ ensembles, so we do not include those two in our analysis. As an example, we show various aspects of the topological charge measured on the $a09m130$ ensemble in Fig.~\ref{fig:Q}. \looseness-1

%% \begin{figure}[tbp]
%% \begin{center}
%% \includegraphics[width=0.495\textwidth]{CorrMap_tmid_ReC2pt%% _P_t08}
%% \includegraphics[width=0.495\textwidth]{CorrMap_tmid_ReV4_U_tsep08}
%% \end{center}
%% \vspace{-0.7cm}
%% \caption{Correlation between topological charge density summed over a 3D-shell on a given timeslice $Q^{\textrm{local}}(R, t) = \sum_{R<|\mathbf{x}_{\textrm{src}}-\mathbf{x}| < R+1} q(\mathbf{x}, t)$ and (left) neutron $C_{\textrm{2pt}}$ calculated with $\gamma_5$ projection and (right) the vector current $V_4$ inserted at $t=4a$. Neutron source is at $t=0$, and sink at $t=8a$\looseness-1. 
%% \label{fig:Qcorr}}
%% \end{figure}

After correlating the topological charge with the neutron 2- and 3-point functions as per Eq.~\eqref{eq:theta-exp}, the CPV phase $\alpha$ arising in the neutron state is obtained by solving
\begin{align}
\label{eq:alpha}
\frac{\textrm{Im} C_{\textrm{2pt}}^{P}(t)}
     {\textrm{Re} C_{\textrm{2pt}}(t)}
\equiv
  \frac{\textrm{Im}\Tr \left[\gamma_5  \frac12 (1+\gamma_4) 
        \langle N(t) \overline N(0) \rangle \right]}
      {\textrm{Re}\Tr \left[ \frac12 (1+\gamma_4) 
       \langle N(t) \overline{N}(0) \rangle \right]}
= \frac{M_N \sin\big(2\alpha(t)\big)}{E_N+M_N\cos\big(2\alpha(t)\big)}\,,
\end{align}
and the electric dipole form-factor $F_3$ is extracted from
\begin{align}
\label{eq:emff}
  \langle N \vert V_\mu(q) \vert N \rangle_{\textrm{CPV}}
    = \overline{u}_N(p') \left[ F_1(q^2)\gamma_\mu 
      + i\frac{F_2(q^2)}{2M_N} \sigma_{\mu\nu} q^\nu 
      - \frac{F_3(q^2)}{2M_N} \sigma_{\mu\nu} q^\nu \gamma_5\right] u_N(p)\,,
\end{align}
where $V_\mu$ is the electromagnetic current, $M_N$ is the neutron mass, $u_N(p)$ is free neutron spinor, $q=p-p'$, and $F_1$ and $F_2$ are the Dirac and Pauli form-factors. The  anapole form factor $F_A$ is irrelevant, assuming $PT$-conservation. Extracting the form factors from the real and imaginary parts of the lattice three-point functions at multiple combinations of the momentum transfer for the same $q^2$ is an over-constrained problem. We solve the equations simultaneously, weighting each by its statistical variance. Details of the extraction of $F_3$ are given in Refs.~\cite{Bhattacharya:2018qat}. The lattice ensembles analyzed and the number of configurations/measurements are listed in Table~\ref{tab:confs}.

\begin{table}
\small
\begin{center}
\begin{tabular}{l|cccc|cc}
  \hline\hline
  Ensemble    & $a$ (fm)  &  $M_\pi$ (MeV) & $L^3\times T$   & $M_\pi L$ & $N_\text{conf}$ & $N_{\rm meas}$  \\\hline
  a15m310 & 0.1510(20) & 320(5) & $16^3\times 48$ & 3.9 & 1920 & 123k\\
  \hline
  a12m310 & 0.1207(11) & 305.3(4) & $24^3\times 64$ & 4.54 & 1013 & 65k \\
  a12m220 & 0.1184(10) & 216.9(2) & $32^3\times 64$ & 4.29 & 1156 & 74k \\
  a12m220L& 0.1189(09) & 217.0(2) & $40^3\times 64$ & 5.36 & 1000 & 128k \\
  \hline
  a09m310 & 0.0888(08) & 312.7(6) & $32^3\times 96$ & 4.50 & 2196 & 140k \\
  a09m220 & 0.0872(07) & 220.3(2) & $48^3\times 96$ & 4.71 &  961 & 123k \\
  a09m130 & 0.0871(06) & 128.2(1) & $64^3\times 96$ & 3.66 & 1289 & 165k \\
  \hline
  a06m135 & 0.0570(01) & 135.5(2) & $96^3\times 192$& 3.70 &  453 & 29k \\
  \hline\hline
\end{tabular}
\end{center}
\vspace{-0.5cm}
\caption{List of MILC HISQ ensembles analyzed for $Q$ and $W_{ggg}$}.\label{tab:confs}
\end{table}

\begin{figure}[tbp]
\begin{center}
\includegraphics[width=0.4\textwidth]{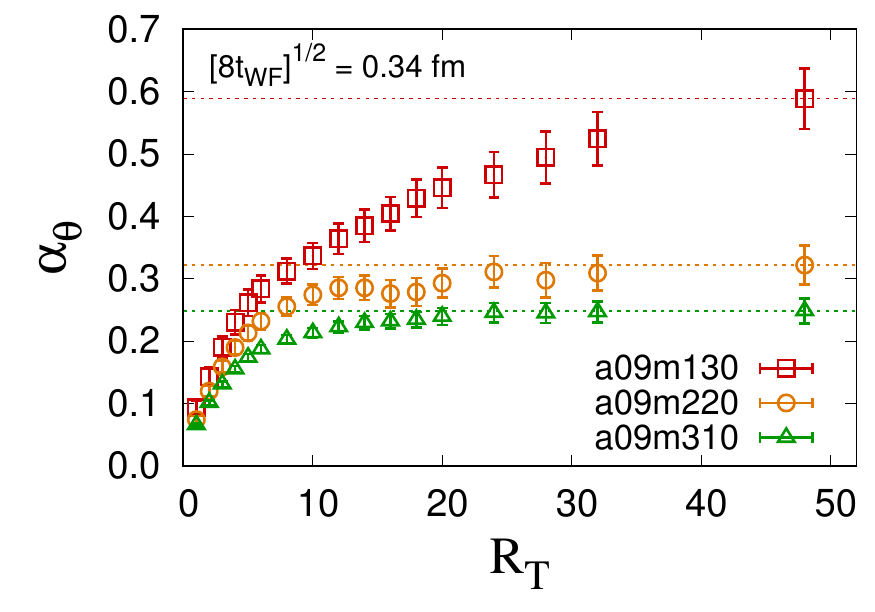} \qquad
\includegraphics[width=0.4\textwidth]{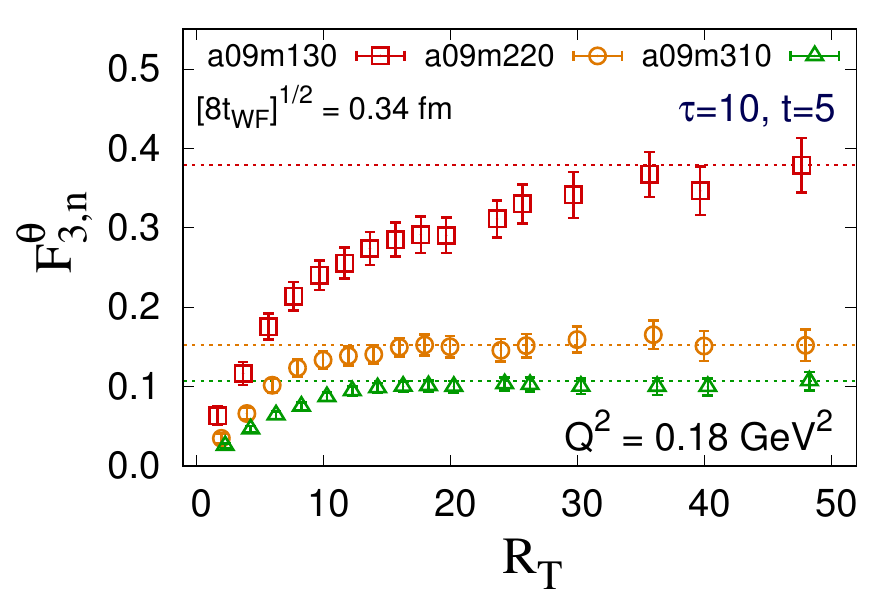}
\end{center}
\vspace{-0.7cm}
\caption{CPV phase $\alpha$ (left) and electric dipole form-factor $F_3$ (right) calculated using the topological charges calculated from the timeslices near the neutron source $\vert t_Q-t_\text{src} \vert \le R_T$ (for $\alpha$) or those from the timeslices near the current insertion $\vert t_Q-t_\text{ins} \vert \le R_T$ (for $F_3$).
\label{fig:TLR}}
\end{figure}

To understand errors, we investigated correlations in the spatial and temporal directions between $Q$ and 2- and 3-point functions. 
%% Fig.~\ref{fig:Qcorr} shows that the correlation between the topological charge density $q(x)$ and the neutron two- and three-point correlators is small when $q(x)$ is far from the neutron source (for two-point correlators) or far from the spatial neutron source position with the current insertion timeslice (for vector currents). We also observe that the correlation remains up to very long spatial distance from the neutron source, while it diminishes relatively faster in the time direction.  
Fig.~\ref{fig:TLR} shows the phase $\alpha_\theta$ and $F_3^\theta$ versus $R_T$, the number of timeslices over which $Q(t)$ is summed about the neutron source or the current insertion time, $\vert t_Q-t_{\text{src, ins}} \vert \le R_T$. For the physical pion mass, where the need for reducing error is the largest, convergence requires $Q(x,t)$ to be summed over almost all $\{x,t\}$. Therefore, we do find any significant advantage to using $Q(x,t)$ summed over a limited volume to reduce errors.

The size of CPV observables $\alpha$ and $F_3$ depend on the parameter $\bar\theta$ used in Eq.~\eqref{eq:theta-exp}. We find that the dependence is linear within errors for the values of $\theta$ used. We, therefore, report results divided by $\theta$, with $\theta=0.2$. We also use the variance reduction technique  (VRT) introduced in \cite{Bhattacharya:2018qat} by calculating 
$\langle O_\textrm{CPV}^{VR} \rangle_{\theta} = \langle O_\textrm{CPV} \vert_{\theta} - c\cdot O_\textrm{CPV} \vert_{\theta=0}\rangle$. Since $\langle O_\textrm{CPV} \rangle_{\theta=0}=0$, adding this does not change the result, but, the error is reduced due to the large correlations. Here $c$ is the coefficient determined following~\cite{Bhattacharya:2018qat}, and it turns out to be $c\approx 1$. This VRT is not useful when $\theta\gtrsim 1$, but becomes crucial when $\theta \ll 1$; for $\theta=0.2$, we find about 25\% reduction in the error of $F_3$.

\begin{figure}[tbp]
\begin{center}
\includegraphics[width=0.45\textwidth]{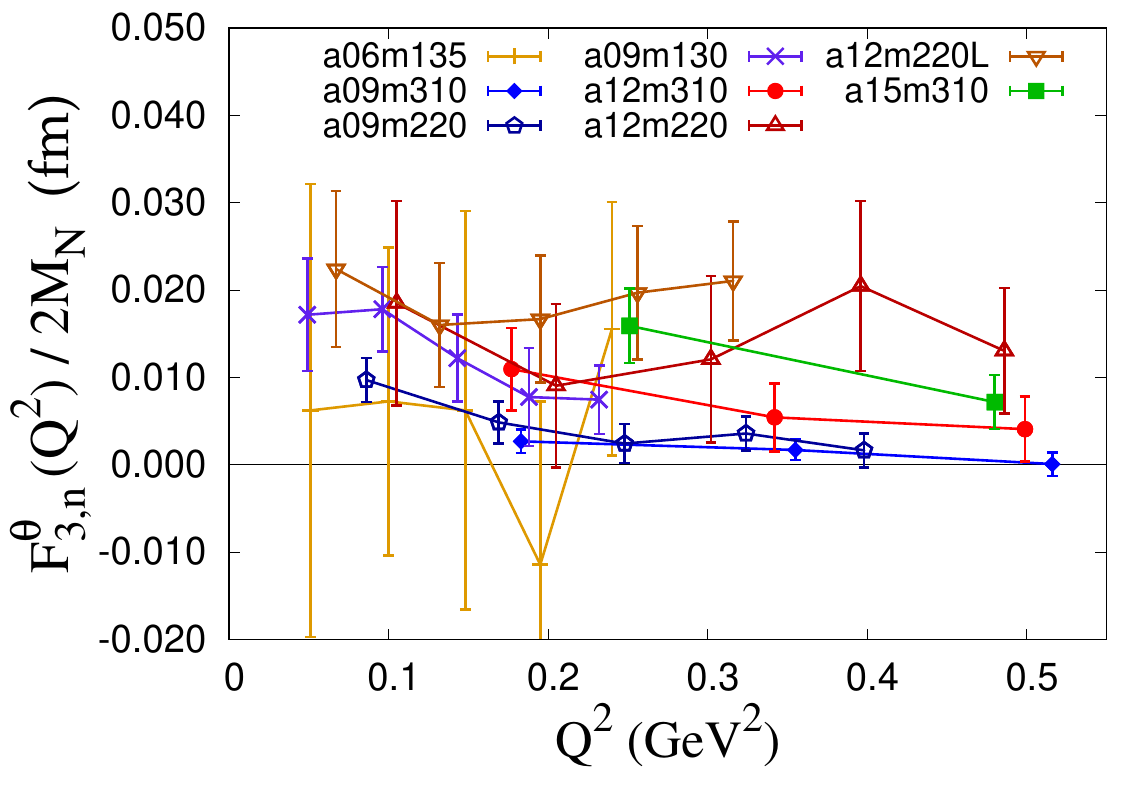} \qquad
\includegraphics[width=0.45\textwidth]{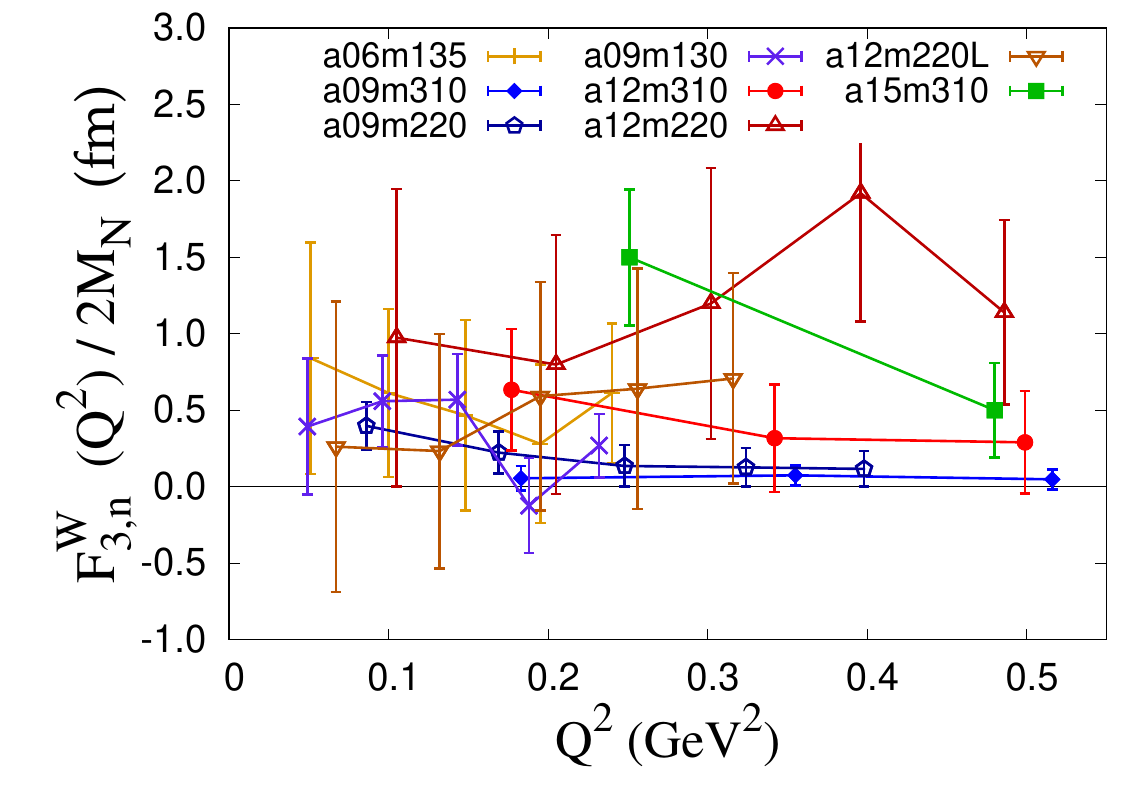}\\
\includegraphics[width=0.45\textwidth]{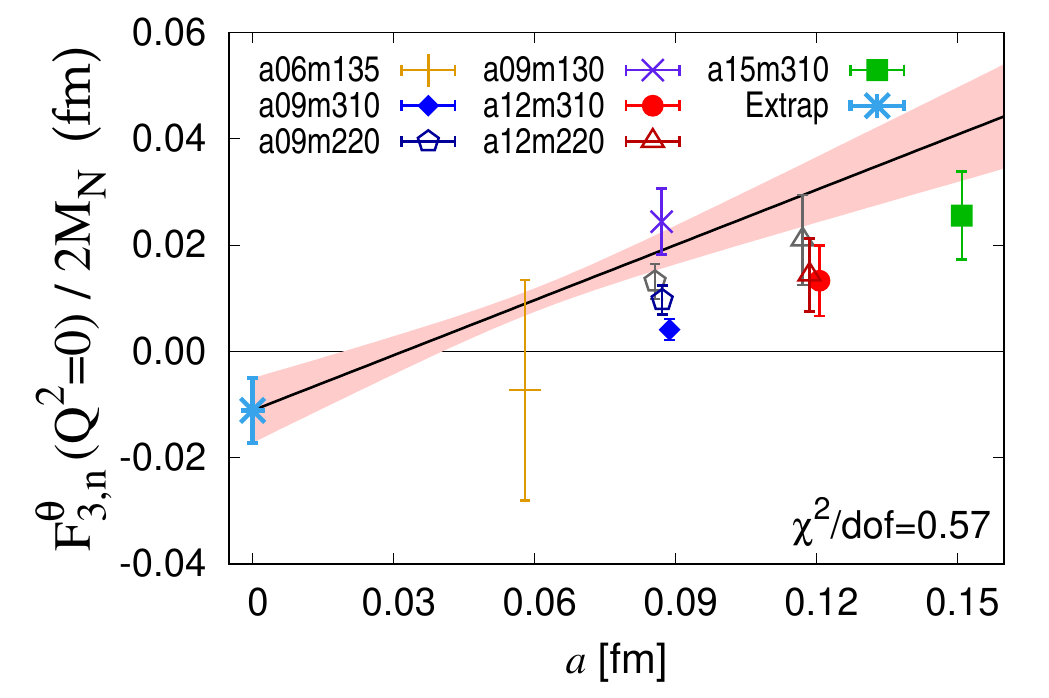} \qquad
\includegraphics[width=0.45\textwidth]{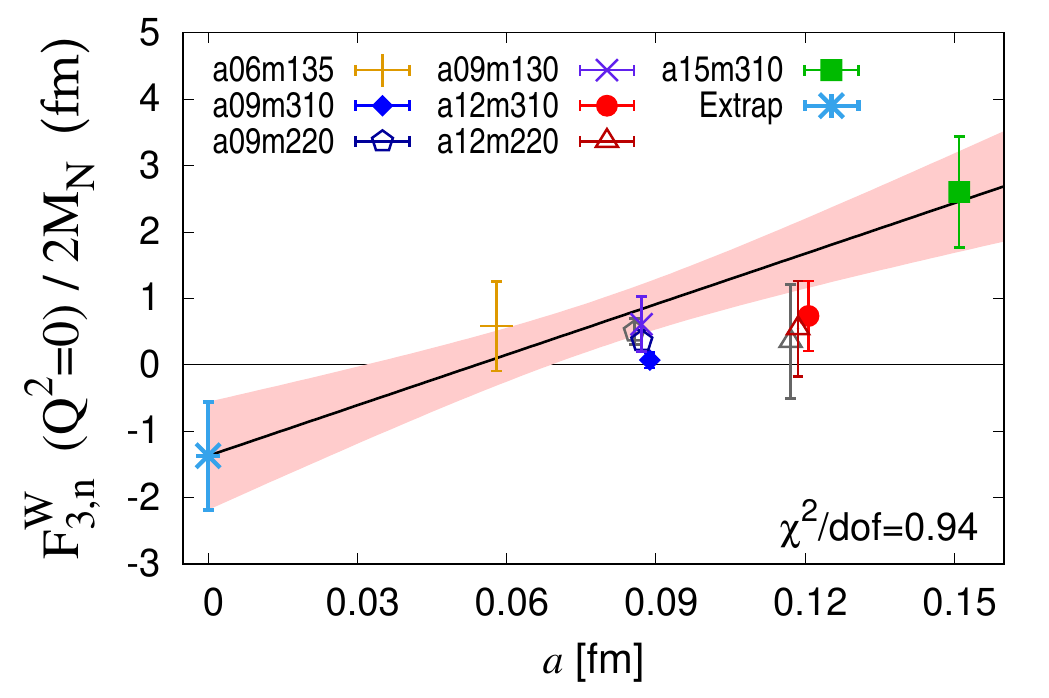}\\
\includegraphics[width=0.45\textwidth]{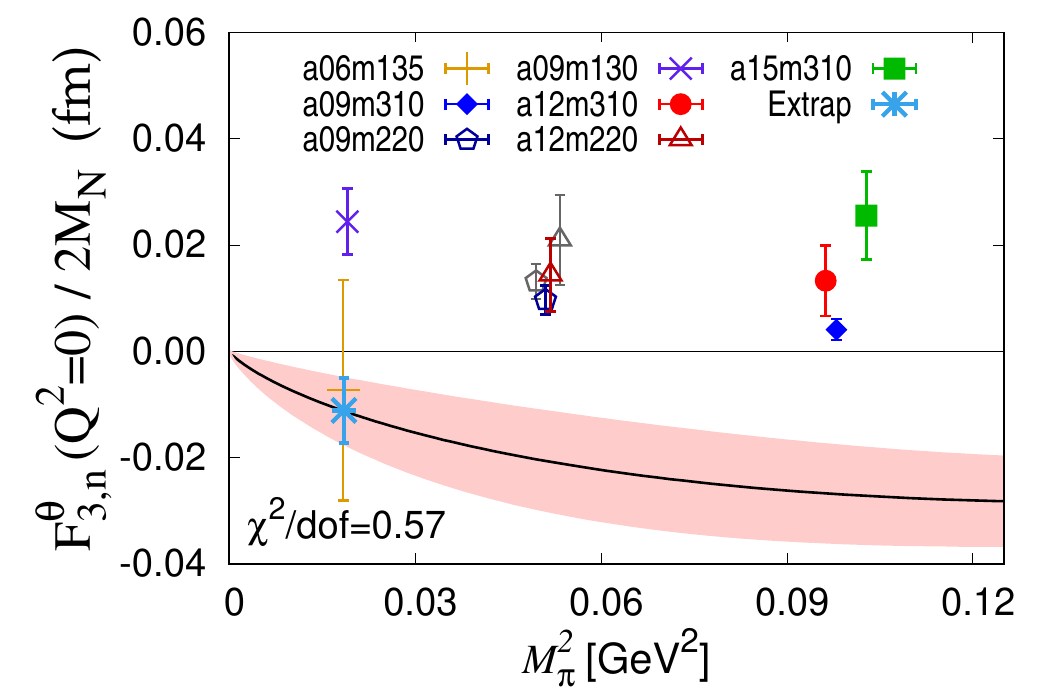} \qquad 
\includegraphics[width=0.45\textwidth]{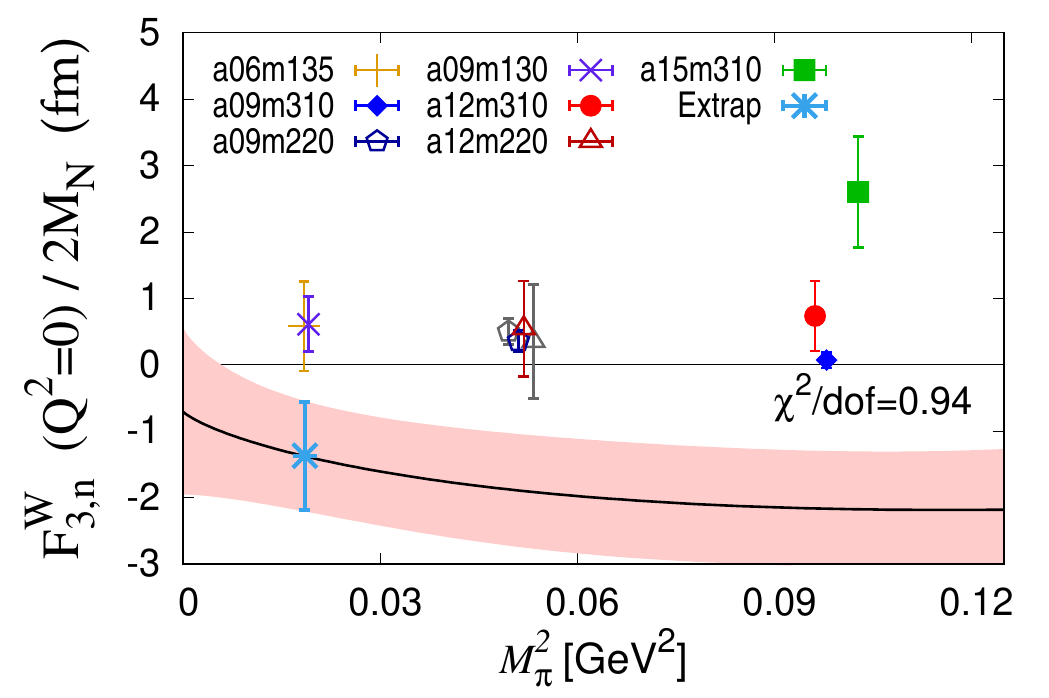}\\
\end{center}
\vspace{-0.7cm}
\caption{Preliminary results on the neutron $F_3/2M_N$ from the QCD $\theta$-term (left) and the $W_{ggg}$ (right). Top row shows $Q^2$-dependence, and bottom two rows show the continuum (middle) and chiral (bottom) extrapolations. Gray data points in the bottom two rows show the $Q^2\rightarrow0$ results using only the $F_3$ data for $Q^2<0.275~\textrm{GeV}^2$. The $F_3$ results from a12m220 and a12m220L ensembles are averaged in the continuum/chiral extrapolation, because no volume dependence is observed.
\label{fig:F3-theta}}
\end{figure}

After calculating $F_3$ for multiple source-sink separations for each $Q^2$, we remove the excited state contamination using the two-state fit ansatz \cite{Bhattacharya:2013ehc} and extrapolate to $Q^2\rightarrow 0$ using a linear-in-$Q^2$ ansatz to obtain ${d_n=|e|F_3(Q^2=0)/2M_N}$ on each ensemble. We repeated the same procedure for $W_{ggg}$. The  chiral-continuum extrapolation is done using the leading chiral term \cite{deVries:2010ah} and linear in $a$:
\begin{align*}
  d_N^\theta = c_1^\theta M_\pi^2 + c_2^\theta M_\pi^2 \log(M_\pi^2/M_{N,\textrm{phys}}^2) + c_3^\theta a\,, \quad
  d_N^W = c_1^W + c_2^W M_\pi^2 + c_3^W M_\pi^2 \log(M_\pi^2/M_{N,\textrm{phys}}^2) + c_4^W a\,.
 \label{eq:fit}
\end{align*}
Preliminary results, presented in Fig.~\ref{fig:F3-theta} in the gradient-flow scheme, are consistent with zero within 2$\sigma$: $d_N^\theta = -0.011(6)\bar\theta |e| \textrm{fm}$ and $d_N^W = -1.4(8)d_w|e| \textrm{fm}$. The extrapolation results fluctuate around zero for different fit ansatze. The results on $M_\pi\approx 310$~MeV ensembles are similar to those from recent lattice calculations \cite{Syritsyn:2019vvt,Dragos:2019oxn}, however, our data show significant discretization corrections.\looseness-1

\section{Neutron EDM from CEDM term}
Since the CEDM operator is a quark bilinear, we use the Schwinger source method (SSM) to include the CEDM interactions by changing the Dirac operator: $D_{clov} \rightarrow D_{clov} + (i/2)\varepsilon \sigma^{\mu\nu}\gamma_5 G_{\mu\nu}$. This is implemented by shifting the Sheikholeslami-Wohlert coefficient, $c_{sw} \rightarrow c_{sw} + 2i\varepsilon\gamma_5$ \cite{Bhattacharya:2016oqm, Bhattacharya:2016rrc}. $F_3$ is then extracted from $\langle N \vert V_\mu(q) \vert N \rangle_{\textrm{CPV}}$ calculated with the modified valence quark propagators. In this study, we ignored the contributions from the disconnected diagrams and the reweighting factor due to the change in the fermion determinant ${\det \left[ D_{clov} + (i/2)\varepsilon \sigma^{\mu\nu}\gamma_5 G_{\mu\nu}\right]}/{\det\left[D_{clov}\right]}$ \cite{Bhattacharya:2016oqm}.

The $\alpha$ is calculated by solving Eq.~\eqref{eq:alpha} and $F_3$ from Eq.~\eqref{eq:emff}. In addition to the CEDM operator, we also calculate $\alpha$ and $F_3$ for  $O_{\gamma_5}\equiv -i\bar{q}\gamma_5 q$, as it mixes with the CEDM operator under renormalization. The ensemble, number of configurations and measurements analyzed are\\
\{(Ens, $N_\text{conf}$, $N_{\rm meas}$)\} = \{($a12m310$, 1012, 130K), ($a12m220L$, 475, 61K), ($a09m310$, 447, 57K)\}.

\begin{figure}
\centering
\begin{minipage}{.5\textwidth}
  \centering
  \includegraphics[width=.8\linewidth]{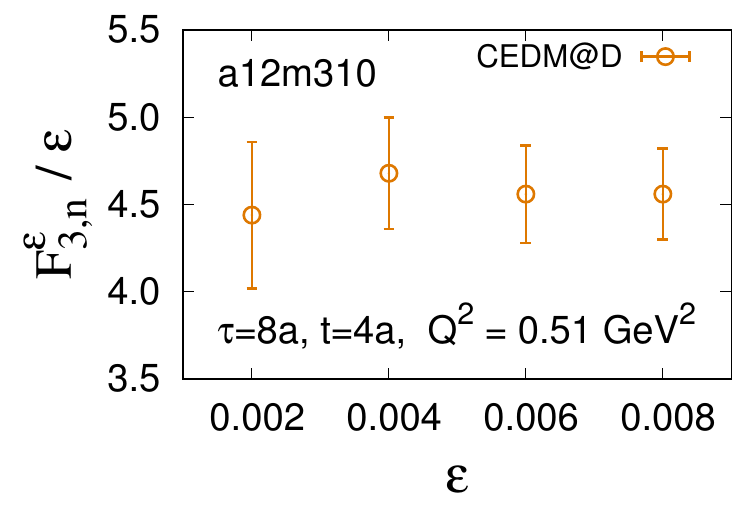}
  \captionof{figure}{Linearity of CEDM $F_3$ in $\varepsilon$.}
  \label{fig:eps-lin}
\end{minipage}%
\begin{minipage}{.5\textwidth}
  \centering
  \includegraphics[width=.8\linewidth]{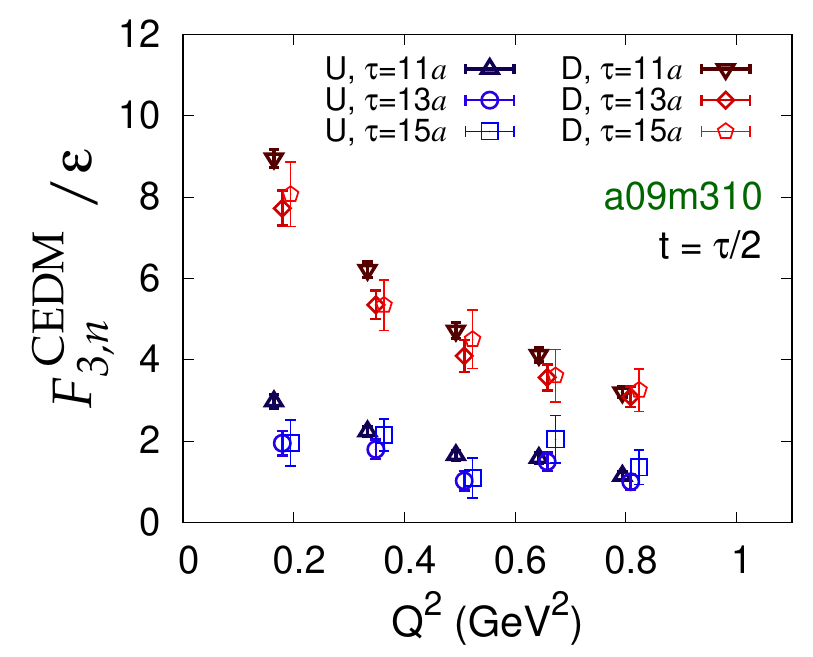}
  \captionof{figure}{CEDM $F_3$ calculated on $a09m310$ at three different source-sink separations.}
  \label{fig:f3_cedm_tsep}
\end{minipage}
\end{figure}

 The SSM assumes $\varepsilon$, simply a parameter, is small so that we can ignore the contributions from $\mathcal{O}(\varepsilon^2)$. This is checked by the linearity of the CPV observables in $\varepsilon$ as shown in Ref.~\cite{Bhattacharya:2016rrc}. Fig.~\ref{fig:eps-lin} shows that $F_3/\varepsilon$ is constant for $\varepsilon \lesssim 0.008$. In general, larger values of $\varepsilon$ give better signal in $F_3$, while for smaller $\varepsilon$ the VRT described above improves the signal significantly. On $a12m310$, for example, without VRT, $F_3^{U, CEDM} /\varepsilon(\tau=8a, t=4a, Q^2=0.50\textrm{GeV}^2)$ is 0.9(2.2) at $\varepsilon=0.002$ and 0.84(55) at $\varepsilon=0.008$. With VRT it becomes 0.98(14) at $\varepsilon=0.002$ and 0.867(82) at $\varepsilon=0.008$.

\begin{figure}[tbp]
\begin{center}
\includegraphics[width=0.43\textwidth]{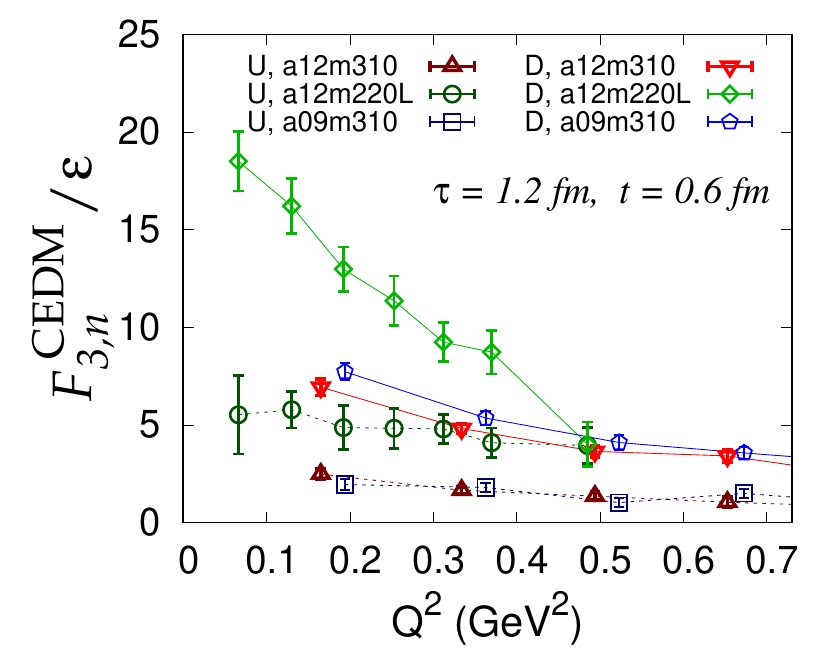} \qquad
\includegraphics[width=0.43\textwidth]{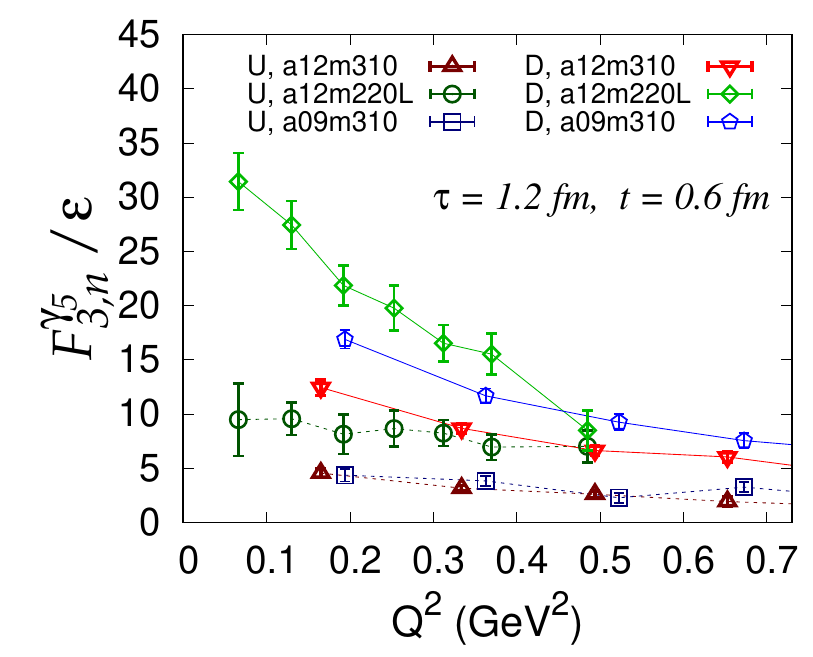}
\end{center}
\vspace{-0.5cm}
\caption{Neutron $F_3/2M_N$ from CEDM as a function of $Q^2$. Note that these are unrenormalized.
\label{fig:F3-cedm}}
\end{figure}

We find relatively small excited state contamination when the source and sink separation ${\tau\gtrsim 1.2}$~fm, as shown in Fig.~\ref{fig:f3_cedm_tsep}. Therefore, we report the $F_3$ obtained at $\tau=1.2$~fm with current inserted in the middle as our final value. The results from all three ensembles are plotted in Fig.~\ref{fig:F3-cedm}. Because the  renormalization factor has not been included, we cannot compare the results from different ensembles, nevertheless, we note the small $a$-dependence between $a09m310$ and $a12m310$ results, and large $M_\pi$-dependence by comparing $a12m310$ and $a12m220L$ results. Similar large $M_\pi$-dependence is also observed in Refs.~\cite{Syritsyn:2019vvt,Abramczyk:2017oxr}. Since we calculate only the connected diagrams, $F_3$ induced by $O_{\gamma_5}$ is a lattice artifact that should disappear in exact chiral symmetry limit~\cite{Guadagnoli:2002nm}.\looseness-1

\section{Conclusion}
Preliminary results of the neutron EDM induced by the QCD $\theta$-term, $W_{ggg}$, and the CEDM interactions, calculated at multiple values of pion masses and lattice spacings are presented. The $a$, $M_\pi$ and the $Q^2$ dependencies of the neutron EDM from $\theta$- and Weinberg-term need further investigation. For the renormalization of the CEDM operator, we are investigating the gradient flow scheme.

%Acknowledgments
\section*{Acknowledgments}
We thank MILC Collaboration for providing the 2+1+1-flavor HISQ lattices. Simulations were carried out on computer facilities at (i) the National Energy Research Scientific Computing Center supported by the U.S. Department of Energy (DOE) Office of Science (OS) under Contract No. DE-AC02-05CH11231; and, (ii) the Oak Ridge Leadership Computing Facility at the Oak Ridge National Laboratory supported by the  DOE OS under Contract No. DE-AC05-00OR22725; (iii) the USQCD Collaboration, which are funded by the DOE OS, (iv) Institutional Computing at Los Alamos National Laboratory. This work was supported by the DOE OS, Office of High Energy Physics under Contract No. 89233218CNA000001, and by the LANL LDRD program.

\bibliographystyle{JHEP}
\bibliography{refs}
\end{document}